\documentclass[11pt,a4paper]{article}
\usepackage{jheppub_kim}

\usepackage{pdflscape}
\usepackage{amsmath}
\usepackage{amssymb}
\usepackage{dcolumn}
\usepackage{bm}
\usepackage{color}
\usepackage{epsfig}
\usepackage{amsfonts}
\usepackage{graphicx}
\usepackage{subfigure}
\usepackage{dcolumn}
 \newcommand{\nn}{\nonumber}

\PassOptionsToPackage{linktocpage}{hyperref}
\def\case#1/#2{\textstyle\frac{#1}{#2}}
\newcommand{\beq}{\begin{equation}}
\newcommand{\eeq}{\end{equation}}
\newcommand{\bea}{\begin{eqnarray}}
\newcommand{\eea}{\end{eqnarray}}

\def\be{\begin{equation}}
\def\ee{\end{equation}}
\def\bea{\begin{eqnarray}}
\def\eea{\end{eqnarray}}

\begin{document}

\title{
Self-Gravitating Spherically Symmetric Solutions in Scalar-Torsion
Theories}

\author[a]{Georgios Kofinas}

\author[b,c]{Eleftherios Papantonopoulos}

\author[b,d]{Emmanuel N. Saridakis}

\affiliation[a]{Research Group of Geometry,
Dynamical Systems and Cosmology,
Department of Information and Communication Systems Engineering,
University of the Aegean, Karlovassi 83200, Samos, Greece}

\affiliation[b]{Physics Division,
National Technical University of Athens, 15780 Zografou Campus,
Athens, Greece}

\affiliation[c]{CERN - Theory Division, CH-1211 Geneva 23,
Switzerland.}

\affiliation[d]{Instituto de F\'{\i}sica, Pontificia Universidad
de Cat\'olica de Valpara\'{\i}so, Casilla 4950, Valpara\'{\i}so,
Chile}

\emailAdd{gkofinas@aegean.gr}

\emailAdd{lpapa@central.ntua.gr}

\emailAdd{Emmanuel$_-$Saridakis@baylor.edu}

%\pacs{04.50.Kd, 98.80.-k, 95.36.+x}

\abstract{ We studied spherically symmetric solutions in
scalar-torsion gravity theories in which a scalar field is coupled
to torsion with a derivative coupling. We obtained the general
field equations from which we extracted a decoupled master
equation, the solution of which leads to the specification of all
other unknown functions. We first obtained an  exact solution
which represents a new  wormhole-like solution dressed with a
regular scalar field. Then, we found large distance linearized
spherically symmetric solutions in which the space
asymptotically is AdS.  }

%\pacs{98.80.-k, 95.36.+x, 04.50.Kd}

\maketitle

%\date{\today}
\newpage

\section{Introduction} \label{Introduction}

One of the main concerns in constructing a physically acceptable
field theory is to avoid unphysical propagating modes, like ghosts.
General Relativity (GR) is one example of a healthy classical theory
of gravity which is based on a unique spacetime metric along with its Levi-Civita
connection,
and it is constructed out of second derivatives in the metric tensor.
The simple form of Hilbert-Einstein action leads to
field equations with no more than  second  derivatives, resulting in
a gravitational theory without ghosts \cite{chaostro}, which moreover, passes
successfully with high accuracy all local observational tests both for
weak and strong gravity \cite{Will:2005va}.

Torsion is an important tensor besides curvature, which can be
used for the construction of gravitational theories. In this case,
one can use as dynamical variables the vielbein and the connection,
and the field equations arise by variation with respect to these
two variables. Soon it was realized that imposing the
teleparallelism condition, it is possible to express equivalently
the standard Einstein gravity
\cite{Unzicker:2005in,TEGR,TEGR22,Hayashi:1979qx,JGPereira,Arcos:2005ec,Maluf:2013gaa,
Pereira:2013qza}
(or even Gauss-Bonnet gravity \cite{Kofinas:2014owa}) by a
different geometry described in terms of torsion. This can be
achieved by employing instead of the Levi-Civita connection the
Weitzenb$\ddot{o}$ck one, which has vanishing curvature but
non-vanishing antisymmetric piece and therefore non-vanishing
torsion, and thus the only dynamical variable that remains is the
vielbein.

In a more general formulation of teleparallelism, an arbitrary connection of
vanishing curvature is assumed, and everything is again expressed
in terms of the torsion, but now both the vielbein and the
connection are dynamical. In this  formulation, the theory is both
Lorentz and diffeomorphism invariant, while the former formulation
lacks local Lorentz invariance. In its teleparallel formulation,
Einstein gravity action consists of the torsion scalar $T,$ which
is a specific combination of quadratic torsion scalars and
contains up to first-order vielbein derivatives. Then, the
variation of this action gives exactly the Einstein gravitational
field equations, and for this reason the theory is termed as
Teleparallel Equivalent of General Relativity (TEGR)
\cite{Unzicker:2005in,TEGR,TEGR22,Hayashi:1979qx,JGPereira,Arcos:2005ec,Maluf:2013gaa,
Pereira:2013qza}.

One interesting question is what kind of theories can be produced
in which the torsion can give sizeable physical effects that can
differentiate these theories from GR. One approach that was put
forward, inspired from the $f(R)$ modifications of GR, is to start
from TEGR instead of GR, and construct $f(T)$ extensions. This new
class of gravitational modification do not coincide with $f(R)$
and proves to have interesting local black hole solutions
\cite{Wang:2011xf,Miao003,Boehmer004,Daouda001,Ferraro:2011ks,Gonzalez:2011dr,
Capozziello:2012zj,
Atazadeh:2012am,Nashed:2013bfa,Paliathanasis:2014iva}, and it
gives variable cosmological models
\cite{Ferraro:2006jd,Ferraro:2008ey,Linder:2010py,Chen:2010va,Dent:2011zz,Bamba:2010wb,
Capozziello006,
Li:2013xea,Ong:2013qja,Kofinas:2014aka,Kofinas:2014daa,Ferraro:2014owa}.

Another approach to probe the effects of torsion to GR, is to
couple the torsion to scalar fields. Self-gravitating scalar
fields were used to study the dynamics of local black hole
solutions of Einstein equations. These studies had resulted to the
well known no-hair theorems \cite{BBMB,bronnikov} and to the way
to evade them
\cite{Martinez:2004nb,Martinez:2005di,Kolyvaris:2009pc,Gonzalez:2013aca,Gonzalez:2014tga}.
In cosmology the scalar fields have been also used to investigate
the features of the early universe
\cite{Sahni:1998at,Uzan:1999ch,Bartolo:1999sq}.
 Apart from
the simple minimal coupling case, the scalar field may be coupled
non-minimally with curvature, since the appearance of terms of the
form $f(\phi)R$ is motivated by many reasons, such as the
variability of the fundamental constants, the Kaluza-Klein
compactification scheme, or the low-energy limit of superstring
theory.

To capture the effects of curvature in the early stages of
cosmological evolution or near the horizon of a black hole,
couplings of scalar fields  directly to curvature through terms
of the form $Rg^{\mu\nu}\partial_{\mu}\phi\,\partial_{\nu}\phi$,
were studied. However, these  derivative coupling terms lead to
field equations containing higher than second derivatives
\cite{Amendola:1993uh}, and therefore ghosts. To  remedy this
pathology one has to couple the scalar fields  to the Einstein
tensor
 \cite{Sushkov:2009hk} and this term $G^{\mu\nu}\partial_{\mu}\phi\,\partial_{\nu}\phi $
is part of
the general
Horndeski Lagrangian  which is known  to result \cite{horny} in a
healthy theory with only second order field equations. These
theories have interesting cosmological implications
\cite{Sushkov:2009hk,Gao:2010vr,Sushkov:2012za,Saridakis:2010mf}
and they give also local black hole solutions
\cite{Kolyvaris:2011fk,Rinaldi:2012vy,Kolyvaris:2013zfa,Cisterna:2014nua,
Charmousis:2014zaa}.

In this work we will discuss first couplings of torsion to scalar
fields of the form $f(\phi)T$\footnote{The cosmological implications
of these type of couplings were studied in
\cite{Geng:2011aj,Geng:2011ka,Xu:2012jf,Otalora:2013tba,Skugoreva:2014ena}}. We will
show that such coupling in four or higher dimensions,  for a
diagonal vielbein, does not give non-trivial spherically symmetric
solutions\footnote{On the contrary, in three dimensions spherically
symmetric solutions were found \cite{Gonzalez:2014pwa}.}.  We will
then introduce a derivative coupling of a scalar field to torsion
of the form $Tg^{\mu\nu}\partial_{\mu}\phi\,\partial_{\nu}\phi$.
This coupling, contrary to its curvature counterpart
$Rg^{\mu\nu}\partial_{\mu}\phi\,\partial_{\nu}\phi$, has the
advantage that it leads to second-order equations of motion. We
will show that such a coupling gives spherically symmetric
solutions which have distinct  features from
their GR counterparts.
Note that the above derivative coupling
between torsion and the scalar field is not the only one that can
be constructed, since other combinations of
$\partial_{\mu}\phi\,\partial_{\nu}\phi$ with quadratic (of even
higher) powers of torsion could appear.

Our work is organized as follows. In Section \ref{model}, after a
brief review of scalar-tensor theories, we set up our theory of a
scalar field coupled to torsion with a derivative coupling and
present the field equations of the gravity-scalar system. In
Section \ref{sola} we consider a spherically symmetric ansatz for
the metric and we reduce the coupled field equations to a
decoupled master equation from which the solutions are obtained.
 In Section \ref{exactsolution} we  discuss an exact  wormhole-like
 solution, while in Section \ref{Largedistance} we discuss large- distance solutions.
Finally, Section \ref{Conclusions} is devoted to our
 conclusions.

\section{Derivative coupling of a scalar field to torsion}
\label{model}

In this section, after a brief review of scalar-tensor theories, and
in particular of a scalar field minimally coupled to gravity, we
will introduce torsion, and we will set up the gravity-scalar
field equations describing the torsion field coupled with a
derivative coupling to a scalar field.

In $D$ dimensions the Einstein-Hilbert action with  a scalar field
  minimally coupled to gravity  is given by
\begin{equation}
S_{min}=\frac{1}{2\kappa_{D}^{2}}\int d^{D}\!x\sqrt{|g|}\,\bar{R}-\int
d^{D}\!x\sqrt{|g|}\,\Big(\frac{1}{2}g^{\mu\nu}
\partial_{\mu}\phi\,\partial_{\nu}\phi+V\Big)\,.
\label{Rminimal}
\end{equation}
The Ricci scalar $\bar{R}$ is constructed from the Christoffel
connection $\Gamma^{\lambda}_{\,\,\,\mu\nu}$. Indices $a,b,...$
will refer to tangent space, while $\mu,\nu,...$ are coordinate
ones. A  cosmological constant $\Lambda=\pm 3l^{-2}$,
where $l$ is the length of the (A)dS space, has been incorporated
in the potential, so the total potential is
$V(\phi)=\frac{\Lambda}{\kappa_{D}^{2}}+V_{1}(\phi)$.
In four dimensions and for a negative cosmological constant this
theory has a local black hole solution known as MTZ solution \cite{Martinez:2004nb} given
by
\begin{equation}
ds^{2}=B(r)\Big[-F(r)dt^{2}+\frac{1}{F(r)}dr^{2}+r^{2}d\sigma^{2}\Big],
\label{bhsol}
\end{equation}
where
\begin{equation}
B(r)=\frac{r(r\!+\!2G_{\!N}\mu )}{(r\!+\!G_{\!N}\mu )^{2}}\,\,\,\,\,,\,\,\,\,\,
F(r)=\frac{r^{ 2}}{l^{2}}-\Big{(}1\!+\!\frac{G_{\!N}\mu }{r}\Big{)}^{2}.
\label{kdd}
\end{equation}
The constant $\mu$ can be identified with the mass of the
black hole and $\kappa{_4}^{2}=8 \pi G_{\!N}$, with $G_{\!N}$ the Newton's
constant. The line element $d\sigma^{2}$ refers to a two-dimensional manifold
of negative constant curvature. The scalar field is given by
\begin{equation}
\phi=\sqrt{\frac{3}{4\pi G_{\!N}}} \,\,\text{arctanh}
\Big(\frac{G_{\!N}\mu}{r+G_{\!N}\mu}\Big)~,
\label{scalarmtz}
\end{equation}
and the potential $V_{1}$ is found to be
\begin{equation}
V_{1}(\phi) = -\frac{3}{4 \pi G_{\!N} l^2}\sinh^2\!\Big(\sqrt{\frac{4\pi
G_{\!N}}{3}}\,\phi\Big)~.
\label{potmtz}
\end{equation}
This is the simplest known hairy black hole solution of a scalar
field minimally coupled to the curvature, which goes to zero at
infinity and it is regular on the horizon. Here we should stress
an important property of this solution. If we switch off the
scalar field we cannot get a black hole solution with fixed mass.
In fact, there is only one integration constant $\mu$, and for
$\phi \rightarrow 0$  the geometry approaches a massless black
hole. This means that for a given mass there are two branches of
different black hole solutions, the one with non-trivial scalar
field given by (\ref{bhsol}), and the vacuum solution (with $\phi
= 0$).

In writing (\ref{Rminimal}) we used the usual  Christoffel
connection which by construction has zero torsion, and thus,
all the gravitational information is incorporated by the curvature
(Riemann) tensor. In the teleparallel formulation an alternative connection
$\omega^{a}_{\,\,\,bc}$
is
adopted, and the basic object is the torsion
\begin{equation}
T^{a}_{\,\,\,bc}=\omega^{a}_{\,\,\,cb}-\omega^{a}_{\,\,\,bc}-e_{b}^{\,\,\,\mu}
e_{c}^{\,\,\,\nu}(e^{a}_{\,\,\,\mu,\nu}-e^{a}_{\,\,\,\nu,\mu})~,
\label{hdo}
\end{equation}
expressed in terms of the arbitrary vielbein $e_a=e^{\,\,\,\mu}_a\partial_\mu$
(with dual $e^a=e^a_{\,\,\, \mu}d x^\mu$) and the connection
$\omega^a_{\,\,\,b}=\omega^a_{\,\,\,b\mu}dx^\mu=\omega^a_{\,\,\,bc}e^c$.
We mention that the torsion as defined in (\ref{hdo}) is a tensor under local
Lorentz transformations and under diffeomorphisms. The vielbein is
assumed to be orthonormal, which implies that the metric is given by
\begin{equation}
g_{\mu\nu}=\eta_{ab}\,e^a_{\,\,\,\mu}\,e^b_{\,\,\,\nu}~,
\end{equation}
where $\eta_{ab}=\text{diag}(-1,1,...1)$ is the Minkowski metric.
The metric is assumed to be compatible with the connection
$\omega^{a}_{\,\,\,bc}$, namely $\eta_{ab|c}=0$, and thus
$\omega_{abc}=-\omega_{bac}$, where $|$ denotes covariant
differentiation with respect to $\omega^{a}_{\,\,\,bc}$. The
teleparallelism condition is imposed by assuming that the
curvature of $\omega^{a}_{\,\,\,bc}$ is zero, i.e. $R^{a}_{\,\,\,
b\mu\nu}\!=\omega^{a}_{\,\,\,b\nu,\mu}-\omega^{a}_{\,\,\,b\mu,\nu}
+\omega^{a}_{\,\,\,c\mu}\omega^{c}_{\,\,\,b\nu}-\omega^{a}_{\,\,\,c\nu}
\omega^{c}_{\,\,\,b\mu}=0$ (note however that the curvature
$\bar{R}^{a}_{\,\,\, b\mu\nu}$ constructed from the Christoffel
connection is not zero). One way to realize this condition is by
considering the Weitzenb{\"{o}}ck connection, which is defined in
terms of the vielbein in all coordinate frames as
$\omega_{\,\,\,\mu\nu}^{\lambda}=e_{a}^{\,\,\,\lambda}e^{a}_{\,\,\,\mu,\nu}$,
and therefore $\omega_{\,\,\,bc}^{a}=0$ in the preferred frame
$e_{a}$ that defines the connection.

Then, the following torsion scalar can be defined as
\begin{eqnarray}
T&=&\frac{1}{4}T^{\mu\nu\lambda}T_{\mu\nu\lambda}+\frac{1}{2}T^{\mu\nu\lambda}
T_{\lambda\nu\mu}-T_{\nu}^{\,\,\,\nu\mu}T^{\lambda}_{\,\,\,\lambda\mu}\label{snn}
\\
&=&S^{\mu\nu\lambda}T_{\mu\nu\lambda}\,, \label{Tquad}
\end{eqnarray}
where the tensor $S^{\mu\nu\lambda}$ is defined by
\begin{equation}
S^{\mu\nu\lambda}=\frac{1}{2}\mathcal{K}^{\nu\lambda\mu}+\frac{1}{2}(g^{\mu\lambda}T_{\rho
}^{\,\,\,\rho\nu}
-g^{\mu\nu}T_{\rho}^{\,\,\,\rho\lambda})=-S^{\mu\lambda\nu}
\label{fyt}
\end{equation}
and
\begin{equation}
\mathcal{K}_{\mu\nu\lambda}=\frac{1}{2}(T_{\lambda\mu\nu}-T_{\nu\lambda\mu}-T_{
\mu\nu\lambda})
=-\mathcal{K}_{\nu\mu\lambda}=\omega_{\mu\nu\lambda}-\Gamma_{\mu\nu\lambda}
\label{lji}
\end{equation}
is the contorsion tensor. In the teleparallel formulation, torsion is used to express the
action (\ref{Rminimal}) in an equivalent form (up to boundary terms) as
\begin{equation}
S_{min}=-\frac{1}{2\kappa_{D}^{2}}\int d^{D}\!x\,e\,T-\int
d^{D}\!x\,e\,\Big(\frac{1}{2}g^{\mu\nu}
\partial_{\mu}\phi\,\partial_{\nu}\phi+V\Big)\,.
\label{Tminimal}
\end{equation}
Since $T$ differs from $\bar{R}$ by boundary terms, namely
$T=-\bar{R}+2T_{\nu\,\,\,\,\,\,;\mu}^{\,\,\nu\mu}$
with $;$ denoting covariant differentiation with respect to the Christoffel connection,
the above
action
(\ref{Tminimal}) indeed gives exactly the same equations with the action (\ref{Rminimal}).
For example, the hairy black hole solution (\ref{bhsol}), (\ref{scalarmtz}),
(\ref{potmtz}) can be
obtained from
the action (\ref{Tminimal}) in four dimensions.

In this work we are interested in finding local spherically
symmetric  solutions in which the presence of the torsion could
leave an explicit signature on the solution, other than from the
trivial one found from the action (\ref{Tminimal}). To achieve
this, the action (\ref{Tminimal}) has to be modified in such a way
as to include a coupling of torsion to the scalar field.

The easiest way to modify the action (\ref{Tminimal}) is to add a
direct coupling of the scalar field to torsion
\begin{equation}
S_{non-min}=-\frac{1}{2\kappa_{D}^{2}}\int d^{D}\!x\,e\,T-\int
d^{D}\!x\,e\,\Big(\frac{1}{2}g^{\mu\nu}
\partial_{\mu}\phi\,\partial_{\nu}\phi+V+\xi f(\phi)T\Big)\,.
\label{Tnonminimal}
\end{equation}
The resulting theory is different from its corresponding curvature counterpart and it
leads
to novel features when applied to a cosmological geometry
\cite{Geng:2011aj,Geng:2011ka,Xu:2012jf,
Otalora:2013tba,Skugoreva:2014ena}.
However, it is not easy to apply the above theory to a spherically
symmetric geometry in four dimensions. The reason is that for a diagonal vierbein the
coupling of the scalar field to torsion gives a new
$(r-\theta)$-equation which acts as a strong constrain to the coupled field equations.
Thus, the arising geometry is trivial and does not give hairy spherically symmetric
solutions.
Note, that in three dimensions the absence of this off-diagonal field equation
allows hairy black hole solutions to be found \cite{Gonzalez:2014pwa}.

Another possible extension of the action (\ref{Tminimal}) is to
include a derivative coupling of the scalar field to torsion. In
particular, we consider such a coupling with the torsion scalar
$T$, namely
\begin{equation}
S=-\frac{1}{2\kappa_{D}^{2}}\int d^{D}\!x\,e\,T-\int d^{D}\!x\,e\,
\Big[\Big(\frac{1}{2}-\xi T\Big)g^{\mu\nu}
\partial_{\mu}\phi\,\partial_{\nu}\phi+V\Big]\,.
\label{Tnonminimal}
\end{equation}
The coupling parameter $\xi$ has dimensions of length square, and therefore,
$\sqrt{|\xi|}$
introduces a new length scale in the theory.

In order to extract the field equations, we need to perform variation with respect to the
vielbein
$e_{a}^{\,\,\,\mu}$ and the scalar field.
Additionally, and more formally, in order to implement the teleparallelism condition
$R_{abcd}=0$,
one adds in the action (\ref{Tnonminimal}) the term $\lambda^{abcd}R_{abcd}$,
where $\lambda^{abcd}$ is a Lagrange multiplier. Thus,
variation  with respect to $\lambda^{abcd}$ naturally provides $R_{abcd}=0$, which can be
considered as the equation of motion for the connection
$\omega^{a}_{\,\,\,bc}$. Variation of the action (\ref{Tnonminimal}) with respect to the
connection $\omega^{a}_{\,\,\,bc}$ gives the equation of motion
for the Lagrange multiplier. Since
\begin{eqnarray}
&&\!\!\!\!-\frac{1}{4}\delta_{e}(eT\phi_{,\mu}\phi^{,\mu})=-eS^{dca}\omega_{bdc}\phi_{,
\rho}\phi^{,\rho}e^{b}_{\,\,\,\mu}\delta e_{a}^{\,\,\,\mu}
-(eS_{a}^{\,\,\,\mu\nu}\phi_{,\kappa}\phi^{,\kappa}\delta e^{a}_{\,\,\,\nu})_{,\mu}
\nn\\
&&\ \ \ \ \ \ \ \ \ \ \ \ \ \ \ \ \ \ \ \ \ \ \
+\left\{\left[(eS_{\kappa}^{\,\,\,\lambda\nu}
e_{b}^{\,\,\,\kappa})_{,\nu}e^{b}_{\,\,\,\mu}
+e\Big(\frac{1}{4}T\delta^{\lambda}_{\mu}-S^{\nu\kappa\lambda}T_{\nu\kappa\mu}\Big)\right]
\phi_{,\rho}\phi^{,\rho}\right.\nonumber\\
&&\left. \ \ \ \ \ \ \ \ \ \ \ \ \ \ \ \ \ \ \ \  \ \ \ \ \, \ \ \
-\frac{1}{2}eT\phi_{,\mu}\phi^{,\lambda}-eS_{\mu}^{\,\,\,\nu\lambda}(\phi_{,\kappa}\phi^{,
\kappa})_{
,\nu}
\right\}
e^{a}_{\,\,\,\lambda}\delta e_{a}^{\,\,\,\mu}\,,
\label{var1}
\end{eqnarray}
it arises
{\small{
\begin{eqnarray}
&&\!\!\!\!\!\!\!\!\!\!\!\!\!\!\!\!\!\!\!\!\!\delta_{e}S=
-\int \!d^{D}\!x\Big(\frac{2}{\kappa_{D}^{2}}\!-\!4\xi\phi_{,\rho}\phi^{,\rho}\Big)
eS^{dca}\omega_{bdc}e^{b}_{\,\,\,\mu}\delta e_{a}^{\,\,\,\mu}
-\int \!d^{D}\!x\left[\Big(\frac{2}{\kappa_{D}^{2}}\!-\!4\xi\phi_{,\rho}\phi^{,\rho}\Big)
eS_{a}^{\,\,\,\mu\nu}\delta e^{a}_{\,\,\,\nu}\right]_{,\mu}
\nn\\
&& \!\! \!\!\!
+\int \!d^{D}\!x\left\{\Big(\frac{2}{\kappa_{D}^{2}}\!-\!4\xi\phi_{,\rho}\phi^{,\rho}\Big)
\left[(eS_{\kappa}^{\,\,\,\lambda\nu} e_{b}^{\,\,\,\kappa})_{,\nu}e^{b}_{\,\,\,\mu}
+e\Big(\frac{1}{4}T\delta^{\lambda}_{\mu}\!-\!S^{\nu\kappa\lambda}T_{\nu\kappa\mu}
\Big)\right]
\right.\nn\\
&&\left. \,\,\,\,\,\,\,\,\,\,\,\,\,\,\,\,
+4\xi\left[\frac{1}{2}eT\phi_{,\mu}\phi^{,\lambda}+eS_{\mu}^{\,\,\,\nu\lambda}(\phi_{,
\kappa}\phi^{,
\kappa})_{,\nu}
\right] \! + \!
e\Big(\frac{1}{2}\phi_{,\rho}\phi^{,\rho}\delta^{\lambda}_{\mu}-\phi_{,\mu}
\phi^{,\lambda}+V\delta^{\lambda}_{\mu}\Big)\right\}
e^{a}_{\,\,\,\lambda}\delta e_{a}^{\,\,\,\mu} , \label{vartot}
\end{eqnarray}}}
$\!\!$where $\phi^{,\mu}=g^{\mu\nu}\phi_{,\nu}$.
Ignoring the boundary term and setting $\delta_{e}S=0$, we obtain
{\small{
\begin{eqnarray}
&&\!\!\!\!\!\!\!\!\!\!\Big(\frac{2}{\kappa_{D}^{2}}\!-\!4\xi\phi_{,\rho}\phi^{,\rho}\Big)
\Big[(eS_{\kappa}^{\,\,\,\lambda\nu} e_{b}^{\,\,\,\kappa})_{,\nu}e^{b}_{\,\,\,\mu}
+e\Big(\frac{1}{4}T\delta^{\lambda}_{\mu}\!-\!S^{\nu\kappa\lambda}T_{\nu\kappa\mu}
\Big)\Big]
+4\xi\Big[\frac{1}{2}eT\phi_{,\mu}\phi^{,\lambda}
\!+\!eS_{\mu}^{\,\,\,\nu\lambda}(\phi_{,\kappa}\phi^{,\kappa})_{,\nu}
\Big]\nn\\
&&\!\!\!\!\!\!\!\!\!\!
+e\Big(\frac{1}{2}\phi_{,\rho}\phi^{,\rho}\delta^{\lambda}_{\mu}-\phi_{,\mu}
\phi^{,\lambda}+V\delta^{\lambda}_{\mu}\Big)
-\Big(\frac{2}{\kappa_{D}^{2}}\!-\!4\xi\phi_{,\rho}\phi^{,\rho}\Big)
eS^{dca}\omega_{bdc}e_{a}^{\,\,\,\lambda}e^{b}_{\,\,\,\mu}=0\,.
\label{equationeomega1}
\end{eqnarray}}}
$\!\!\!$We mention that the above expressions contain also the connection $\omega_{abc}$.
Assuming for simplicity the Weitzenb$\ddot{o}$ck connection, that
is setting $\omega_{abc}=0$, which is a solution of the equation
of motion $R_{abcd}(\omega^{a}_{\,\,\,bc})=0$, we acquire the field equations containing
only the vielbein, namely
\begin{eqnarray}
&&\!\!\!\!\!\!\!\!\!\!\!\!\!\!\!\!\!\!\!\!\!\!\!\!\!
\Big(\frac{2}{\kappa_{D}^{2}}\!-\!4\xi\phi_{,\rho}\phi^{,\rho}\Big)
\left[(eS_{\kappa}^{\,\,\,\lambda\nu} e_{b}^{\,\,\,\kappa})_{,\nu}e^{b}_{\,\,\,\mu}
+e\Big(\frac{1}{4}T\delta^{\lambda}_{\mu}\!-\!S^{\nu\kappa\lambda}T_{\nu\kappa\mu}
\Big)\right]
\nn\\
&&\!\!\!\!\!\!\!\!\!\!\!\!\!\!\!\!\!\!\!\!\!\!\!\!\!
+4\xi\left[\frac{1}{2}eT\phi_{,\mu}\phi^{,\lambda}
 + eS_{\mu}^{\,\,\,\nu\lambda}(\phi_{,\kappa}\phi^{,\kappa})_{,\nu}
\right]
+e\Big(\frac{1}{2}\phi_{,\rho}\phi^{,\rho}\delta^{\lambda}_{\mu}-\phi_{,\mu}
\phi^{,\lambda}+V\delta^{\lambda}_{\mu}\Big)=0\,.
\label{equationeomega1}
\end{eqnarray}
Additionally, variation of the action (\ref{Tnonminimal}) with respect to the scalar field
$\phi$ gives
\begin{equation}
\delta_{\phi}S=\int\!d^{D}\!x
\left\{\big[e(1-2\xi T)\phi^{,\mu}\big]_{,\mu}-e\frac{dV}{d\phi}\right\}\delta\phi
-\int\!d^{D}\!x\big[e(1-2\xi T)\phi^{,\mu}\delta\phi\big]_{,\mu}\,,
\label{variationphi}
\end{equation}
from where the scalar-field equation of motion is obtained as
\begin{equation}
\big[e(1-2\xi T)\phi^{,\mu}\big]_{,\mu}-e\frac{dV}{d\phi}=0\,.
\label{equationphi}
\end{equation}

Summarizing, equations (\ref{equationeomega1}) are the gravity
field equations, while equation (\ref{equationphi}) is the
Klein-Gordon one for the scalar field. Observe that $T$ being
a scalar can be identified  with the Hubble parameter in a
cosmological Friedmann-Robertson-Walker (FRW) background as
$T=G_{00}=6H^2$. This implies that the derivative coupling
 $Tg^{\mu\nu}\partial_{\mu}\phi\,\partial_{\nu}\phi$ on a
 FRW background will give the same cosmological evolution as the
derivative coupling of the scalar field to Einstein tensor
$G^{\mu\nu}\partial_{\mu}\phi\,\partial_{\nu}\phi$. However,
general cosmological perturbations will differentiate the two
models because they will excite also the spatial components of the
Einstein tensor.

In the next section we will study local spherically symmetric
solutions of the field equations (\ref{equationeomega1}) and
(\ref{equationphi}) and we expect to find solutions with features
different than  the local black hole solutions of theories with a
scalar field coupled to Einstein tensor with a derivative coupling
\cite{Kolyvaris:2011fk,Rinaldi:2012vy,Kolyvaris:2013zfa,Cisterna:2014nua,
Charmousis:2014zaa}.

\section{Spherically symmetric solutions}
\label{sola}

In this section we look for spherically symmetric solutions of the
theory with a derivative coupling of the torsion to the scalar
field in four dimensions. We consider the general spherically
symmetric metric ansatz of the form
\begin{align}
ds^{2}=-N(r)^2 dt^2+K(r)^{-2}dr^2 +R(r)^2 d\Omega^2\,,
\label{sphermetric}
\end{align}
where $d\Omega^2 =d\theta^2+\sin^2\!\theta\,d\varphi^2$ is the
two-dimensional sphere, and $N(r)$, $K(r)$ and $R(r)$ are three
unknown functions. Note that we do not impose the relation
$N(r)=K(r)$ keeping the metric ansatz as general as possible.
We could have chosen the gauge with $R(r)=r$, but we find it more convenient
not to impose a particular gauge and to let also the function $R(r)$ free.
We consider the following diagonal vierbein which gives rise to the above metric
\begin{equation}
e^a_{\,\,\,\mu}
={\text{diag}}\left(N(r),K(r)^{-1},R(r),R(r)\sin\!\theta\right)~,
\label{vierb1}
\end{equation}
which is adequate to capture the main features of our theory.

Inserting the vierbein (\ref{vierb1}) into the definition of the
torsion scalar (\ref{snn}) we obtain
\begin{equation}
T(r) = -2K^2\frac{R'}{R} \Big(\frac{R'}{R}+\frac{2N'}{N}\Big)
\label{Torsionsca}~,
\end{equation}
where primes denote derivatives with respect to $r$ and we
denote $\kappa^{2}=\kappa_{4}^{2}=8 \pi G_{\!N}$. Also the field
equations (\ref{equationeomega1}) become
\begin{eqnarray}
&&
\frac{1}{K^{2}}\Big(\phi'^{2} + \frac{2V}{K^{2}}\Big)+2\Big(\frac{1}{\kappa^{2}K^{2}}
 - 2\xi\phi'^{2}\Big)
\Big(\frac{R'^{2}}{R^{2}} + \frac{2R''}{R} + 2\frac{R'}{R}\frac{K'}{K} -
\frac{1}{K^{2}R^{2}}\Big)\nonumber\\
&&\ \ \ \ \ \ \ \ \ \ \ \ \ \ \ \ \ \ \ \ \ \ \ \ \ \ \ \ \ \ \ \ \ \ \ \ \ \ \ \ \ \ \ \
\ \ \ \ \
\ \ \
-16\xi\phi'\frac{R'}{R}\Big(\frac{K'}{K}\phi' + \phi''\Big)=0~,
\label{1}
\\
&&
\frac{1}{K^{2}}\Big(\phi'^{2}- \frac{2V}{K^{2}}\Big)
 +2\Big(2\xi\phi'^{2} - \frac{1}{\kappa^{2}K^{2}}\Big)\left[\frac{R'}{R}
\Big(\frac{R'}{R} + \frac{2N'}{N}\Big) - \frac{1}{K^{2}R^{2}}\right]
\nonumber\\
&& \ \ \ \ \ \ \ \ \ \ \ \ \ \ \ \ \ \ \ \ \ \ \ \ \ \ \ \ \ \ \ \ \ \ \ \ \  \ \ \ \ \ \
\ \ \, \ \ \ \ \ \ \ \
+8\xi\phi'^{2}\frac{R'}{R}\Big(\frac{R'}{R} +
\frac{2N'}{N}\Big)=0~,\label{2}
\end{eqnarray}
\begin{eqnarray}
&&
\frac{1}{K^{2}}\Big(\phi'^{2} + \frac{2V}{K^{2}}\Big)+2\Big(\frac{1}{\kappa^{2}K^{2}}
 - 2\xi\phi'^{2}\Big)
\left[\frac{N'}{N}\Big(\frac{R'}{R} + \frac{K'}{K}\Big) + \frac{R'}{R}\frac{K'}{K} +
\frac{N''}{N}
 + \frac{R''}{R}\right]\nonumber\\
&&\ \ \ \ \ \ \ \ \ \ \ \ \ \ \ \ \ \ \ \ \ \ \ \ \ \ \ \ \ \ \ \ \ \ \ \ \ \ \ \ \ \ \ \
\ \ \ \
\ \
-8\xi\phi'\Big(\frac{R'}{R} +\frac{N'}{N}\Big)\Big(\frac{K'}{K}\phi' +
\phi''\Big)=0~,\label{3}
\\
&& \phi'\Big(\frac{K'}{K}\phi' + \phi''\Big)=0~. \label{4}
\end{eqnarray}
Equations (\ref{1})-(\ref{4}) are accordingly the
$(\lambda,\mu)=tt,rr,\theta\theta,\theta r$ components of the
system (\ref{equationeomega1}), while the other components vanish.
The Klein-Gordon equation of the scalar field (\ref{equationphi}) becomes
\begin{equation}
\left\{KNR^{2}\phi'\left[1+4\xi
K^{2}\frac{R'}{R}\Big(\frac{R'}{R}\!+\!\frac{2N'}{N}\Big)\right]\right\}'
-\frac{NR^{2}}{K}\frac{dV}{d\phi}=0~. \label{eqmsc}
\end{equation}

The off-diagonal equation (\ref{4}) is a new feature of four or
higher-dimensional teleparallel gravity and it does not appear  in
the corresponding curvature-based equations of motion. It rises
from the specific metric  ansatz (\ref{sphermetric}) and the
vielbein (\ref{vierb1}) in spite that they are diagonal. Its
appearance constrains the system of field equations making them
more difficult to be solved. Its physical significance is that it
connects directly the scalar field to the metric function $K(r)$
of (\ref{sphermetric}).

Equations (\ref{1})-(\ref{eqmsc}) are  invariant under
$r$-reparametrizations, in the sense that when $r\rightarrow
\tilde{r}(r)$ and $K\rightarrow K\frac{d\tilde {r}}{dr}$,
$N\rightarrow N$, $R\rightarrow R$, $\phi\rightarrow \phi$, the
equations remain form invariant. This implies that one of these
equations is expected to form the constraint of the theory, as will indeed be shown
in the following. More specifically  we have four functions, namely
$N(r),K(r),\phi(r),V(r)$ (the function $R(r)$ is not considered in the
enumeration because it corresponds to the choice of the radial gauge)
and there are five equations of motion
(\ref{1})-(\ref{eqmsc}), but one of them is the constraint.
So finally, we end up with four true equations for four unknowns.
This is an interesting feature because the solution of the system for the
diagonal vielbein will be basically unique without the need of any extra
assumptions. On the contrary, in the curvature-based theories, for
the same ansatz (\ref{sphermetric}) the absence of the
off-diagonal equation (\ref{4}) provides three equations for the
four unknowns. Therefore, one function has to be fixed arbitrarily,
making the solutions quite sensitive to initial assumptions.

Assuming a non-trivial profile for the scalar field,  equation
(\ref{4}) is integrated to give
\begin{equation}
\phi'=\frac{\nu}{K}\,,
\label{int1}
\end{equation}
where $\nu$ is an integration constant with dimensions of inverse
length square. This is expected to characterize the charge of the scalar field.
In section  \ref{exactsolution} we find an exact solution where  the constant $\nu$ is
related to
the  derivative coupling $\xi$
which determines the strength of the interaction of torsion to matter.
However, in section  \ref{Largedistance}, in the linearized solution found, these two
parameters
remain unrelated.

 Using
(\ref{int1}) equations (\ref{1})-(\ref{3}) and (\ref{eqmsc}),
along with the replacement of
$\frac{R''}{R}+\frac{R'}{R}\frac{K'}{K}$ from (\ref{1}) into
(\ref{3}), acquire the following form
\begin{eqnarray}
\frac{2R''}{R}+\frac{R'^{2}}{R^{2}}+2\frac{R'}{R}\frac{K'}{K}-\frac{1}{K^{2}R^{2}}+\frac{
\kappa^{2}}
{2(1\!-\!2\xi\kappa^{2}\nu^{2})}\frac{2V\!+\!\nu^{2}}{K^{2}}=0~,
\label{111}
\end{eqnarray}
\begin{eqnarray}
\frac{1\!-\!6\xi\kappa^{2}\nu^{2}}{\kappa^{2}}\frac{R'}{R}\Big(\frac{R'}{R}+\frac{2N'}{N}
\Big)
+\Big(V-\frac{\nu^{2}}{2}-\frac{1\!-\!2\xi\kappa^{2}\nu^{2}}{\kappa^{2}}\frac{1}{R^{2}}
\Big)\frac{1}
{K^{2}}=0~,
\label{222}
\\
\frac{N''}{N}+\frac{R''}{R}+\frac{N'}{N}\Big(\frac{R'}{R}\!+\!\frac{K'}{K}\Big)
+\frac{R'}{R}\frac{K'}{K}+\frac{\kappa^{2}}{2(1\!-\!2\xi\kappa^{2}\nu^{2})}\frac{
2V\!+\!\nu^{2}}{K^{
2}}=0~,
\label{333}\\
\left\{NR^{2}\left[1+4\xi
K^{2}\frac{R'}{R}\Big(\frac{R'}{R}\!+\!\frac{2N'}{N}\Big)\right]\right\}'
-\frac{NR^{2}}{\nu K}\frac{dV}{d\phi}=0\,.
\label{555}
\end{eqnarray}

The above system of equations is quite complicated and we aim,
after performing a series of transformations to obtain a single,
decoupled, master equation from where we can get the behaviour of
the system. To do so, we consider   as independent argument the
variable $\phi$ instead of the coordinate $r$ and we find
 for any $f(r)$ that $f'=\frac{\nu}{K}\dot{f}$,
$f''=\frac{\nu^{2}}{K^{2}}(\ddot{f}\!-\!
\dot{f}\frac{\dot{K}}{K})$, where $\dot{f}=\frac{df}{d\phi}$,
$\ddot{f}=\frac{d^{2}f}{d\phi^{2}}$. Then the system of field
equations (\ref{111})-(\ref{555}) becomes
\begin{eqnarray}
\frac{\ddot{R}}{R}+\frac{\dot{R}^{2}}{2R^{2}}
-\frac{1}{2\nu^{2}R^{2}}+\frac{\kappa^{2}(2V\!+\!\nu^{2})}
{4\nu^{2}(1\!-\!2\xi\kappa^{2}\nu^{2})}=0~,
\label{1b}\\
\frac{\dot{R}}{R}\Big(\frac{\dot{R}}{R}+\frac{2\dot{N}}{N}\Big)
+\frac{\kappa^{2}}{\nu^{2}(1\!-\!6\xi\kappa^{2}\nu^{2})}
\Big(V-\frac{\nu^{2}}{2}-\frac{1\!-\!2\xi\kappa^{2}\nu^{2}}{\kappa^{2}}\frac{1}{R^{2}}
\Big)=0~,
\label{2b}\\
\frac{\ddot{N}}{N}+\frac{\dot{N}}{N}\frac{\dot{R}}{R}
-\frac{\dot{R}^{2}}{2R^{2}}+\frac{1}{2\nu^{2}R^{2}}
+\frac{\kappa^{2}(2V\!+\!\nu^{2})}{4\nu^{2}(1\!-\!2\xi\kappa^{2}\nu^{2})}=0~,
\label{3b}\\
\left\{NR^{2}\left[1+4\xi\nu^{2}\frac{\dot{R}}{R}\left(\frac{\dot{R}}{R}\!+\!\frac{2\dot{N
} }{N}\right)
\right]\right\}^{^{\cdot}}-\frac{NR^{2}}{\nu^{2}}\dot{V}=0\,.
\label{5b}
\end{eqnarray}
Replacing (\ref{3b}) by the sum of (\ref{1b}) and (\ref{3b}) and
defining
\begin{eqnarray}
&&x=\ln{R}\label{xsmall}~,\\
&&y=\frac{\dot{R}}{R}\label{ysmall}~,\\
&&z=\frac{(RN^{2})^{^{\cdot}}}{RN^{2}}\,,\label{zsmall}
\end{eqnarray}
equations (\ref{1b})-(\ref{5b}) become
\begin{eqnarray}
\dot{y}+\frac{3}{2}y^{2}
-\frac{e^{-2x}}{2\nu^{2}}+\frac{\kappa^{2}(2V\!+\!\nu^{2})}
{4\nu^{2}(1\!-\!2\xi\kappa^{2}\nu^{2})}=0\,,\label{1ccf}\\
yz=\frac{\kappa^{2}}{\nu^{2}(1\!-\!6\xi\kappa^{2}\nu^{2})}
\Big(\frac{1\!-\!2\xi\kappa^{2}\nu^{2}}{\kappa^{2}}e^{-2x}+\frac{\nu^{2}}{2}-V\Big)\equiv{
F}\,,
\label{2ccf}\\
\dot{z}+\frac{1}{2}z^{2}+\frac{e^{-2x}}{2\nu^{2}}+
\frac{3\kappa^{2}(2V\!+\!\nu^{2})}{4\nu^{2}(1\!-\!2\xi\kappa^{2}\nu^{2})}=0\,,
\label{3ccf}\\
(yz)^{\cdot}-\frac{1}{4\xi\nu^{4}}\dot{V}+\frac{1}{2}(3y\!+\!z)\Big(\frac{1}{4\xi\nu^{2}}
\!+\!yz\Big)=0\,.
\label{5ccf}
\end{eqnarray}
Thus, from (\ref{1ccf}), (\ref{3ccf}) we have
\begin{equation}
(yz)^{\cdot}+\frac{1}{2}\left[yz\!+\!\frac{\kappa^{2}(2V\!+\!\nu^{2})}{2\nu^{2}
(1\!-\!2\xi\kappa^{2}
\nu^{2})}\right]
(3y\!+\!z)+\frac{1}{2\nu^{2}}e^{-2x}(y\!-\!z)=0\,,
\label{hdk}
\end{equation}
while (\ref{2ccf}) gives
\begin{equation}
\dot{V}=-2\frac{1\!-\!2\xi\kappa^{2}\nu^{2}}{\kappa^{2}}e^{-2x}y-\frac{\nu^{2}
(1\!-\!6\xi\kappa^{2}\nu^{2})}{\kappa^{2}}(yz)^{\cdot}\,.
\label{wig}
\end{equation}
Multiplying equation (\ref{wig}) by
$-\frac{\kappa^{2}}{\nu^{2}(1-2\xi\kappa^{2}\nu^{2})}$  and adding
this to equation (\ref{hdk}) we get a new equation, in which when
$V$ is replaced from (\ref{2ccf}), we obtain equation
(\ref{5ccf}). This reveals that equation (\ref{2ccf}) is a
constraint, and thus, equation (\ref{5ccf}) is redundant.

Note that the system of equations (\ref{1})-(\ref{3}) for vanishing scalar field $\phi=0$
and a
cosmological
constant as the potential $V$ has the standard Schwarzschild-(A)dS solution. However, in
the
equivalent system
(\ref{1ccf})-(\ref{5ccf}) this limit has been lost, since setting $\nu=0$ in this system
we get
inconsistency.

Defining the new variables
\begin{eqnarray}
&&Y\equiv y^{2}\label{Yy2}\\
&&Z\equiv z^{2}\label{Zz2}~,
\end{eqnarray}
and using the field equations (\ref{1ccf})-(\ref{3ccf}),
we get a differential equation for  $Y(x)$ as
\begin{equation}
2\frac{d^{2}Y}{dx^{2}}-\frac{1}{Y}
\Big(\frac{dY}{dx}\Big)^{2}+2\Big(2-\frac{\eta\nu^{2}}{Y}\Big)\frac{dY}{dx}
+3Y+2\eta\nu^{2} -12\frac{\eta^{2}}{\tilde{\eta}^{2}}Y
\frac{\frac{dY}{dx}+3Y-\frac{2}{3\nu^{2}}e^{-2x}}
{\frac{dY}{dx}+3Y+2\eta\nu^{2}}=0\,. \label{jsitext}
\end{equation}
Equation (\ref{jsitext}) consists the master equation which we were
looking for, while the technical details for deriving this equation
are given in the Appendix A. The quantities $\eta,\tilde{\eta}$ are
defined by equations (\ref{hud}), (\ref{owk}).  The strategy which we will follow is to
find a
solution of
this equation and then use the other decoupled equations to specify
the metric, the scalar field profile and the potential.

In particular, finding the function $Y$ from   (\ref{jsitext})
allows us to use
equations (\ref{2ccc}), (\ref{las}) in order to find $Z$ as
\begin{equation}
Z=\frac{\tilde{\eta}^{2}}{4\eta^{2}Y}\Big(\frac{dY}{dx}+3Y+2\eta\nu^{2}\Big)^{2}\,.
\label{gij}
\end{equation}
Moreover, combining equations (\ref{bdj}), (\ref{koi}), (\ref{las}), we find the potential
$V$ as
\begin{equation}
V=\frac{1}{2\nu^{2}\eta}e^{-2x}-\frac{\nu^{2}}{2}-\frac{1}{2\eta}\Big(\frac{dY}{dx}
+3Y\Big)\,.
\label{wij}
\end{equation}
Additionally, from equation (\ref{Yy2}) and the definitions (\ref{xsmall}), (\ref{ysmall})
we have
\begin{equation}
\Big(\frac{dx}{d\phi}\Big)^{2}=Y(x)\,,
\label{dxdphiv}
\end{equation}
which can be integrated to provide $\phi(R)$. Then, we can use equations (\ref{zsmall}),
(\ref{Zz2})
and (\ref{dxdphiv}) to obtain
\begin{equation}
\left[\frac{d\ln(RN^2)}{dx}\right]^{2}=\frac{Z(x)}{Y(x)}\,,
\label{dlnRN21}
\end{equation}
the integration of which provides $N(R)$. Finally, using (\ref{int1}), (\ref{xsmall}) and
(\ref{dxdphiv}) we can find
\begin{equation}
K^{-2}dr^{2}=\frac{dR^{2}}{\nu^{2}R^{2}Y}~, \label{oem}
\end{equation}
and therefore  the metric (\ref{sphermetric}) acquires the form
\begin{equation}
ds^{2}=-N^{2}dt^{2}+\frac{dR^{2}}{\nu^{2}R^{2}Y}+R^{2}d\Omega^{2}\,.
\label{jsb}
\end{equation}

Although we have managed to extract from the  initial coupled
system of field equations the  master equation (\ref{jsitext})
plus decoupled equations, the general solution of (\ref{jsitext})
is still difficult to be determined analytically. In the next two
sections we will give two solutions of the theory and discuss
their physical significance. The first one is a special exact solution of
the field equations where the potential turns out to have a simple
form. The second one is obtained from an asymptotic solution of
the master equation (\ref{jsitext}).

\section{A new class of wormhole-like solutions}
\label{exactsolution}

In this section we will find a special exact solution of the above system of equations,
and in particular written in the form  (\ref{1ccc})-(\ref{3ccc}). Consider the
potential
\begin{eqnarray} \label{poteff}
 V(x)=\alpha e^{-2x}+\beta~,
\end{eqnarray}
where $\alpha$ and $\beta$ are constants. Then, from equations
(\ref{Y}) and (\ref{Z}) we obtain
\begin{eqnarray}
&&Y(x)=c_{1}e^{-3x}+\left(\frac{1}{\nu^{2}}-2\eta
\alpha\right)e^{-2x}
-\frac{\eta}{3}(\nu^{2}+2\beta)~,\label{Yx}\\
&&Z(x)=c_{2}e^{-x}-\left(\frac{1}{\nu^{2}}-6\eta
\alpha\right)e^{-2x} -3\eta(\nu^{2}+2\beta)\,, \label{Zx}
\end{eqnarray}
where $c_{1},c_{2}$ are integration constants. Choosing
$c_{1}=c_{2}=0$ we find
\begin{eqnarray}
&& Y(x)=Ae^{-2x}-B\label{Yxx}~,\\
&& Z(x)=Ce^{-2x}-D\label{Zxx}\,,
\end{eqnarray}
where
\begin{eqnarray}
&&A=\frac{1}{\nu^{2}}-2\eta \alpha~,\nn\\
&&B=\frac{\eta}{3}(\nu^{2}+2\beta)~,\nn\\
&&C=6\eta \alpha-\frac{1}{\nu^{2}}~,\nn\\
&&D=9B\,. \label{keb}
\end{eqnarray}
Thus, inserting (\ref{Yxx}), (\ref{Zxx}) in (\ref{2ccc}), we find
that this equation is satisfied by choosing
\begin{eqnarray}
&&\!\!\!\!\!\!\!\!\!\!\!\!\!\!\!\!\!\!\!\!\!\!\!\!\!\!\!\!\!\!\!\!\!\!\!\!\!\!\!\!\!\!
\!\!\!\!\!\!\!\!\!\!\!\!\!\!\!\!\! \!\text{Case
I\,\,\,:}\,\,\,\,\,\,\,\,\,\alpha=\frac{5}{8\kappa^{2}}
\,\,\,,\,\,\,\beta=\nu^{2}=\frac{1}{8\kappa^{2}\xi}~,\nonumber
\\
&& \!\!\!\!\!\!\!\!\!\!\!\!\!\!\!\!\!\!\!\!\!\!\!\!\!\!\!\!\!\!
A=\frac{4\kappa^{2}\xi}{3}
\,\,\,,\,\,\,B=\frac{2\kappa^{2}}{3}\,\,\,,\,\,\,C=12\kappa^{2}\xi
\,\,\,,\,\,\,D=6\kappa^{2}~, \label{alge1}
\end{eqnarray}
or
\begin{eqnarray}
&&
\!\!\!\!\!\!\!\!\!\!\!\!\!\!\!\!\!\!\!\!\!\!\!\!\!\!\!\!\!\!\!\!\!\!\!\!\!\!\!\!\!\!
\!\!\!\!\!\!\!\!\!\!\!\!\!\! \! \text{Case
II\,\,\,:}\,\,\,\,\,\,\,\,\,\alpha=\frac{1}{2\kappa^{2}}
\,\,\,,\,\,\,\beta=\nu^{2}=\frac{1}{5\kappa^{2}\xi}~,\nonumber
\\
&&
\!\!\!\!\!\!\!\!\!\!\!\!\!\!\!\!\!\!\!\!\!\!\!\!\!A=\frac{5\kappa^{2}\xi}{6}
\,\,\,,\,\,\,B=\frac{5\kappa^{2}}{6}\,\,\,,\,\,\,C=\frac{15\kappa^{2}\xi}{2}
\,\,\,,\,\,\,D=\frac{15\kappa^{2}}{2}~. \label{alge2}
\end{eqnarray}

Since $Y,Z\geq 0$, it is implied that $Ae^{-2x}-B\geq 0$,
$Ce^{-2x}-D\geq 0$, and therefore $A,C> 0$, which requires
$\xi>0$. Observe that in both cases, because the constant $\beta$
of the potential (\ref{poteff}) is proportional to $1/ \xi$, it is
scaled as $l^{-2}$ and therefore acts as a positive cosmological
constant. Also note that the charge of the scalar field $\nu$ is
proportional to $\sqrt{1/\xi}$. Moreover, we deduce that $R\leq
\sqrt{\frac{A}{B}}, \sqrt{\frac{C}{D}}$. This means that for the
case (\ref{alge1}) it is $R\leq \sqrt{2\xi}$, while for the case
(\ref{alge2}) it is $R\leq \sqrt{\xi}$.

Now, knowing the solution for $Y(x)$ from (\ref{Yxx}), we can use
equation (\ref{dxdphiv}) to find the scalar field as
\begin{equation}
\phi(R)=\phi_1+\frac{\epsilon_{1}}{\sqrt{B}}\arctan\left(\!\sqrt{\frac{A}{B}
\frac{1}{R^{2}}-1}\,\right)\,, \label{dxdphi2}
\end{equation}
where $\epsilon_{1}=\pm 1$ and $\phi_{1}$ is an integration
constant. It proves convenient to redefine the field $\phi$,
absorbing the integration constant $\phi_{1}$ and the index
$\epsilon_{1}$. In particular, setting
$\tilde{\phi}(R)=\epsilon_{1}[\phi(R)-\phi_{1}]$  we result to
\begin{equation}
\tilde{\phi}(R)=\frac{1}{\sqrt{B}}\arctan\left(\!\sqrt{\frac{A}{B}
\frac{1}{R^{2}}-1}\,\right). \label{uie}
\end{equation}
 Thus, for the case I it is
\begin{equation}
\tilde{\phi}(R)=\sqrt{\frac{3}{2}}\frac{1}{\kappa}\arctan\left(\!\sqrt{\frac{2\xi}{R^{2}}
-1}\,\right)\,, \label{uie1}
\end{equation}
while for the case II we have
\begin{equation}
\tilde{\phi}(R)=\sqrt{\frac{6}{5}}\frac{1}{\kappa}\arctan\left(\!\sqrt{\frac{\xi}{R^{2}}
-1}\,\right)\,. \label{klc}
\end{equation}

The potential $V$ can be now reconstructed by inverting (\ref{dxdphi2}), namely
\begin{equation}
V(\phi)=\frac{\alpha
B}{A}\tan^{2}\left[\sqrt{B}(\phi-\phi_{1})\right]+\frac{\alpha
B}{A}+\beta\,, \label{vdo1}
\end{equation}
or
\begin{equation}
V(\tilde\phi)=\frac{\alpha
B}{A}\tan^{2}\big(\sqrt{B}\,\tilde{\phi}\,\big)+\frac{\alpha
B}{A}+\beta\,. \label{vdo}
\end{equation}
Hence, for the case I we have
\begin{equation}
V(\tilde{\phi})=\frac{5}{16\xi\kappa^{2}}\tan^{2}\left(\sqrt{\frac{2}{3}}\,\kappa\,
\tilde{\phi}\right)+\frac{7}{16\xi\kappa^{2}}\,, \label{jdi}
\end{equation}
while for the case II it is
\begin{equation}
V(\tilde{\phi})=\frac{1}{2\xi\kappa^{2}}\tan^{2}\left(\sqrt{\frac{5}{6}}\,\kappa\,
\tilde{\phi}\right)+\frac{7}{10\xi\kappa^{2}}\,. \label{jdd}
\end{equation}
The dependence of $V$ to $R$ for the case I is
\begin{equation}
V(R)=\frac{1}{8\xi\kappa^{2}}+\frac{5}{8\kappa^{2}R^{2}}\,, \label{era}
\end{equation}
while for the case II
\begin{equation}
V(R)=\frac{1}{5\xi\kappa^{2}}+\frac{1}{2\kappa^{2}R^{2}}\,. \label{ema}
\end{equation}

Note that the solutions for the scalar field and the potential
have analogous forms with the corresponding solutions of the
scalar field and potential of the MTZ solutions (relations
(\ref{scalarmtz}) and (\ref{potmtz}) respectively). However, while the
cosmological constant $\Lambda=\kappa^{2} V(\tilde{\phi}=0)$ of
the MTZ solution is an arbitrary parameter, here, the corresponding cosmological
constant is given for case I by $\Lambda=\frac{7}{16\xi}$ and for case II by
$\Lambda=\frac{7}{10\xi}$. Thus, the gravity-scalar field equations have fixed
this constant to depend on the coupling of torsion to the scalar field. Not only this,
but also the integration constant $\nu$ of the scalar field is determined from the
coupling $\xi$ through Eqs. (\ref{alge1}), (\ref{alge2}). This feature also
occurs to the solutions of scalar-tensor theories, where usually
the scalar charge of a local hairy black hole solution is
related to the other scales of the theory like mass or charge,
unless a symmetry is spontaneously broken, like conformal symmetry
\cite{Kolyvaris:2009pc}.

The lapse function can be calculated  from equation (\ref{wdd}).
Using the parameter values from (\ref{keb}), (\ref{alge1}),
(\ref{alge2}), we find
\begin{equation}
\frac{d\ln(RN^2)}{dx}=3\epsilon\,, \label{fse}
\end{equation}
which has as a solution
\begin{equation}
N(R)=\mathcal{C}R^{\frac{1}{2}(3\epsilon-1)}\,, \label{kwo}
\end{equation}
with $\epsilon=\pm 1$ and $\mathcal{C}$ an integration constant.
Therefore,
\begin{equation}
N=\mathcal{C}R
\,\,\,\,\,\,\,\text{for}\,\,\,\epsilon=1\,\,\,\,\,\,\,\,\,\,\,\,\,\,\,\,\,\,
\text{or}\,\,\,\,\,\,\,\,\,\,\,\,\,\,\,\,\,\,
N=\frac{\mathcal{C}}{R^{2}}
\,\,\,\,\,\,\,\text{for}\,\,\,\epsilon=-1\,. \label{ksk}
\end{equation}

Finally, inserting the above relations into  (\ref{jsb}), we find
the metric
\begin{equation}
ds^{2}=-\mathcal{C}^{2}R^{3\epsilon-1}dt^{2}+\frac{dR^{2}}{\nu^{2}(A-BR^{2})}+R^{2}
d\Omega^{2}\,\,\,
\,
,\,\,\,\,\epsilon=\pm 1. \label{jsl}
\end{equation}
Thus, for the case I it is
\begin{equation}
ds^{2}=-\mathcal{C}^{2}R^{3\epsilon-1}dt^{2}+\frac{12\xi\,
dR^{2}}{2\xi-R^{2}}+R^{2}d\Omega^{2}\,, \label{jvo}
\end{equation}
while for the case II we have
\begin{equation}
ds^{2}=-\mathcal{C}^{2}R^{3\epsilon-1}dt^{2}+\frac{6\xi\,
dR^{2}}{\xi-R^{2}}+R^{2}d\Omega^{2}\,. \label{whj}
\end{equation}

These metrics, together with $\tilde{\phi}(R)$ from (\ref{uie})
and $V(\tilde{\phi})$ from (\ref{vdo}) form the exact spherically
symmetric  solutions that we have found. The solutions contain the
two couplings $\kappa,\xi$ and one integration constant
$\mathcal{C}$. Note that when the metric function in (\ref{jvo})
or (\ref{whj}) has a singularity, i.e. at the maximum value
$R^{2}=2\xi$ or $R^{2}=\xi$, the accompanied scalar field
$\tilde{\phi}$ vanishes, $\tilde{\phi}=0$. Moreover, at the origin
$R=0$, the scalar field is still finite. Therefore, the scalar
field is regular everywhere. However, the potential $V$ becomes
infinite at the origin $R=0$, as it is seen from (\ref{era}),
(\ref{ema}).  Finally, the Ricci and the
Kretchmann scalars diverge at the origin but are finite elsewhere.
For example, for the metric (\ref{whj}) and $\epsilon=1$ it is
$R_{\mu\nu\kappa\lambda}R^{\mu\nu\kappa\lambda}=\frac{2R^{4}+2\xi
R^{2}+9\xi^{2}}{3\xi^{2}R^{4}}$.

Defining $b(R)=\nu^{2}BR^{3}+(1-\nu^{2}A)R$ the metric (\ref{jsl})
can be written in the form
\begin{equation}
ds^{2}=-\mathcal{C}^{2}R^{3\epsilon-1}dt^{2}+\frac{dR^{2}}{1-b(R)/R}+R^{2}d\Omega^{2}~,
\label{worm}
\end{equation}
where for Case I it is $b(R)=\frac{1}{12\xi}R^{3}+\frac{5}{6}R$,
while for Case II it is $b(R)=\frac{1}{6\xi}R^{3}+\frac{5}{6}R$.
The lapse function $N(R)$ is always regular,
therefore, the metric (\ref{worm}) does not represent a black hole.
It has the form of a wormhole \cite{Visser:1995cc,Lemos:2003jb}
 with ``throat'' at
$R=\sqrt{A/B}$. The difference is, however, that here the solution
extends inside the ``throat'' up to the origin $R=0$, while in a
usual wormhole the solution extends outside the ``throat'' up to the
``mouth'' or up to infinity. The behaviour of this wormhole-like
solution is controlled by the length  scale $\xi$ and this
solution merits more investigation.

\section{Spherically symmetric solutions at large distances }
\label{Largedistance}

In this section we will solve the master equation (\ref{jsitext})
perturbatively. We will assume that for $x\rightarrow \infty$,
where $x$ is related to the radial distance as (\ref{xsmall}), the
function $Y(x)$ has the expansion $Y(x)=Y_0+Y_1(x)+...$, where $|Y_{1}(x)|\ll |Y_{0}|$.
So, we will obtain large distance linearized solutions.

At zero order (\ref{jsitext}) gives
\begin{equation}
Y_0 =\frac{2\eta\nu^{2}}{3(2\sigma\!-\!1)}\,, \label{kdo}
\end{equation}
where we have introduced the constant
\begin{equation}
\sigma\equiv\frac{\varepsilon\,\eta}{\tilde{\eta}} \label{kjq}
\end{equation}
with $\varepsilon=\pm 1$. Since, due to (\ref{Yy2}), it is $Y>0$, we
have  $Y_0 >0$ and therefore $\eta(2\sigma-1)>0$. Obviously the
exponential factor in (\ref{jsitext}) is suppressed compared to the
asymptotic value $Y_{0}$. To continue with the linearized
correction $Y_{1}$ we assume additionally that $|Y_1(x)| \ll |Y_0
+\frac{2\eta\nu^{2}}{3}|$. The master equation (\ref{jsitext})
at first order, after substituting the zeroth order value
(\ref{kdo}), gives
\begin{equation}
\frac{d^{2}Y_1}{dx^{2}}+
\left[4(1\!-\!\sigma)+(1\!-\!2\sigma)\frac{e^{-2x}}{6\eta\nu^{4}}\right]\frac{dY_1}{dx}
+3(1-2\sigma)\left[1+(1\!-\!2\sigma)\frac{e^{-2x}}{6\eta\nu^{4}}\right]Y_1
+\frac{2\sigma e^{-2x}}{3\nu^{2}}=0\,. \label{din}
\end{equation}
Assuming further that $R\gg
\frac{1}{2\nu^{2}}\sqrt{\frac{|1-2\sigma|}{6|\eta||1-\sigma|}}$,
equation (\ref{din}) becomes
\begin{equation}
\frac{d^{2}Y_1} {dx^{2}}+4(1\!-\!\sigma)\frac{dY_1}{dx}
+3(1\!-\!2\sigma)Y_1+\frac{2\sigma e^{-2x}}{3\nu^{2}}=0\,.
\label{don}
\end{equation}
The general solution of (\ref{don}) is given by
\begin{equation}
Y_1=c_{\lambda}e^{\lambda x}+c_{\mu}e^{\mu
x}+\frac{2\sigma}{3\nu^{2}(1\!-\!2\sigma)} e^{-2x}\,, \label{kei}
\end{equation}
where $c_{\lambda}, c_{\mu}$ are integration constants and
\begin{equation}
\lambda=2(\sigma\!-\!1)\!-\!\sqrt{1\!-\!2\sigma\!+\!4\sigma^{2}}<0\,\,\,\,\,\,,\,\,\,\,\,
\,\,
\mu=2(\sigma\!-\!1)\!+\!\sqrt{1\!-\!2\sigma\!+\!4\sigma^{2}}\,,
\label{woi}
\end{equation}
with $\mu>\lambda$ being either positive or negative.
Depending on the possible parameter values we have two cases:

\begin{itemize}

\item

{\underline{Case A}}

For $\sigma<\frac{1}{2}$ we have $\eta<0$. Then the only
consistent case is for $\varepsilon=-1$,
$\xi\kappa^{2}\nu^{2}>\frac{1}{2}$ (thus the coupling $\xi$ must
be positive), and finally  $\sigma<-\frac{3}{2}$. The values of
$\lambda,\mu$ are $\lambda<-2$, $-3/2<\mu<0$, and therefore the total
linearized solution becomes
\begin{eqnarray}
Y=\frac{2|\eta|\nu^{2}}{3(1\!-\!2\sigma)}+c_{m}e^{-m
x}+\frac{2\sigma}{3\nu^{2}(1\!-\!2\sigma)}
e^{-2x}+c_{\ell}e^{-\ell x} \label{jck}
\end{eqnarray}
with
\begin{eqnarray}
&&0<m=2(1\!-\!\sigma)\!-\!\sqrt{1\!-\!2\sigma\!+\!4\sigma^{2}}<\frac{3}{2}\nn\\
&&\ell=2(1\!-\!\sigma)\!+\!\sqrt{1\!-\!2\sigma\!+\!4\sigma^{2}}>2\,,
\end{eqnarray}

or in terms of the radial distance $R$
\begin{eqnarray}
Y=\frac{2|\eta|\nu^{2}}{3(1\!-\!2\sigma)}+c_{m}R^{-m}+\frac{2\sigma}{3\nu^{2}
(1\!-\!2\sigma)}
R^{-2}+c_{\ell}R^{-\ell}\,, \label{eck}
\end{eqnarray}
where $c_{\ell},c_{m}$ are integration constants. According to the previous
assumptions of linearization, the above solution is valid provided that
$\Big|c_{m}R^{2-m}+\frac{2\sigma}{3\nu^{2}(1-2\sigma)}
+\frac{c_{\ell}}{R^{\ell-2}}\Big|\ll\frac{2|\eta|\nu^{2}}{3(1-2\sigma)}R^{2},
\frac{4|\eta|\nu^{2}|\sigma|}{3(1-2\sigma)}R^{2}$.
These inequalities mean that $R$ should be larger than some minimum distance, which
decreases as $c_
{m}$
decreases.

Using the above solution of the master equation  and following the
procedure described in  Section \ref{sola}, the metric (\ref{jsb})
is found to be
\begin{equation}
ds^{2}=-N^{2}dt^{2}+\frac{dR^{2}}{\frac{2|\eta|\nu^{4}}{3(1-2\sigma)}R^{2}+\texttt{c}_{m}
R^{2-m}
+\frac{2\sigma}{3(1-2\sigma)}+\frac{\texttt{c}_{\ell}}{R^{\ell-2}}}+R^{2}d\Omega^{2}\,,
\label{duk}
\end{equation}
where $\texttt{c}_{m}=\nu^{2}c_{m}$,
$\texttt{c}_{\ell}=\nu^{2}c_{\ell}$ are redefined integration
constants. The lapse  metric function $N$ can be obtained from the
equations (\ref{gij}) and (\ref{dlnRN21}) from which we obtain
\begin{equation}
\frac{d\ln(RN^2)}{dx}=\frac{\zeta}{2\sigma
Y}\Big(\frac{dY}{dx}+3Y+2\eta\nu^{2}\Big)\,, \label{wdd}
\end{equation}
where $\zeta=\pm 1$ is a sign symbol. Integrating equation
(\ref{wdd}) we extract the lapse function as
\begin{equation}
N^{2}(R)=\frac{c}{R}\left[R^{3}\,Y\!(R)
\,e^{2\eta\nu^{2}\!J(R)}\right]^{\frac{\zeta}{2\sigma}}\,,
\label{djk}
\end{equation}
where
\begin{equation}
J(R)=\int\frac{dR}{R\,Y\!(R)}=\frac{\nu^{2}}{2}\int\frac{du}
{\frac{2|\eta|\nu^{4}}{3(1-2\sigma)}u+\texttt{c}_{m}u^{1-\frac{m}{2}}
+\frac{2\sigma}{3(1-2\sigma)}+\frac{\texttt{c}_{\ell}}{u^{\frac{\ell}{2}-1}}}\Bigg{|}_{
u=R^{2}}
\label{hjh}
\end{equation}
and $c>0$ is an integration constant.

The  potential can be
determined after the insertion of the solution (\ref{eck}) into
(\ref{wij}) to find
\begin{equation}
V=\frac{(1\!+\!2\sigma)\nu^{2}}{2(1\!-\!2\sigma)}-\frac{3\!-\!m}{2\eta\nu^{2}}\frac{
\texttt{c}_{m}}{
R^{m}}
+\frac{3\!-\!8\sigma}{6(1\!-\!2\sigma)\eta\nu^{2}}\frac{1}{R^{2}}
+\frac{3\!-\!\ell}{2\eta\nu^{2}}\frac{\texttt{c}_{\ell}}{R^{\ell}}\,.
\label{ekv}
\end{equation}
Finally the scalar field configuration can be determined inserting
the solution (\ref{eck}) into (\ref{dxdphiv}) to obtain
\begin{equation}
\phi=\phi_{1}+\epsilon_{1}\nu\int{\frac{dR}{\sqrt{\frac{2|\eta|\nu^{4}}{3(1-2\sigma)}R^{2}
+\texttt{
c}_{m}R^{2-m}
+\frac{2\sigma}{3(1-2\sigma)}+\frac{\texttt{c}_{\ell}}{R^{\ell-2}}}}}\,,
\label{wnd}
\end{equation}
where $\epsilon_{1}$ is another $\pm 1$ sign and $\phi_{1}$ is an integration constant.
%Although the above integral cannot be performed analytically, however, the leading
asymptotic
%behaviour of the scalar field is logarithmic, namely
%\begin{equation}
%\phi\approx\phi_{1}+\frac{\epsilon_{1}}{\nu}\sqrt{\frac{3(1\!-\!2\sigma)}{2|\eta|}}\,\ln{
%\!R}\,.
%\label{wnds}
%\end{equation}
 Going back to the potential
(\ref{ekv}), we can find the next-to-leading order correction of $V$
as a function of the scalar field, namely
\begin{equation}
V(\tilde{\phi})\approx\frac{(1\!+\!2\sigma)\nu^{2}}{2(1\!-\!2\sigma)}
-\frac{(3\!-\!m)\texttt{c}_{m}}{2\eta\nu^{2}}
e^{-m\nu\sqrt{\frac{2|\eta|}{3(1-2\sigma)}}\,\tilde{\phi}}\,,
\label{dwe}
\end{equation}
where $\tilde{\phi}=\epsilon_{1}\big(\phi-\phi_{1}\big)$.

The integral (\ref{hjh}) cannot be performed analytically,
however, the leading asymptotic behaviour is found to be $J(R)\approx
\frac{3(1-2\sigma)}{2|\eta|\nu^{2}}\ln{\!R}$,
therefore, the leading asymptotic behaviour of the lapse function is
\begin{equation}
N^{2}(R)\approx
c\left[\frac{2|\eta|\nu^{2}}{3(1\!-\!2\sigma)}\right]^{\frac{\zeta}{2\sigma}}
\frac{1}{R^{1-3\zeta}}\,.
\label{gfc}
\end{equation}
Of course, the exact integral (\ref{hjh}) gives more information at even smaller
distances.
The asymptotic form of the metric (\ref{jsb}) becomes
\begin{equation}
ds_{\infty}^{2}=-c\left[\frac{2|\eta|\nu^{2}}{3(1\!-\!2\sigma)}\right]^{\frac{\zeta}{
2\sigma}}
\frac{1}{R^{1-3\zeta}}dt^{2}+\frac{3(1\!-\!2\sigma)}{2|\eta|\nu^{4}}\frac{dR^{2}}{R^{2}}
+R^{2}d\Omega^{2}\,.
\label{ecl}
\end{equation}
Therefore, for the branch with $\zeta=1$, rescaling the time $t$ to
$\tilde{t}=\frac{\sqrt{c}}{|\nu|}
\big[\frac{2|\eta|\nu^{2}}{3(1-2\sigma)}\big]^{\frac{\zeta}{4\sigma}-\frac{1}{2}}\,t$, we
find the
asymptotic metric
\begin{equation}
ds_{\infty}^{2}=-\frac{2|\eta|\nu^{4}}{3(1\!-\!2\sigma)}R^{2}d\tilde{t}^{2}
+\frac{3(1\!-\!2\sigma)}{2|\eta|\nu^{4}}\frac{dR^{2}}{R^{2}}+R^{2}d\Omega^{2}\,.
\label{ecd}
\end{equation}

The resulted metric (\ref{ecd}) indicates that the space is
asymptotically AdS and some interesting features appear. The
length scale introduced here is given by $l_{\text{eff}}^{-2}
=\frac{2|\eta|\nu^{4}}{3(1-2\sigma)}$ and the corresponding
effective cosmological constant is $\Lambda_{\text{eff}}=-3
l_{\text{eff}}^{-2}$. On the other hand, the potential $V$
of the solution asymptotically has a different negative
constant given by Eq. (\ref{ekv}) for $R\rightarrow
\infty$. Therefore, the  constant coming from the potential of the
action and the effective cosmological constant of the
asymptotically AdS spacetime can have different values, and this
is due to the fact that the scalar field acquires a non-trivial
profile at infinity. In this way it introduces another energy
scale asymptotically. Note that the ratio of
$\Lambda_{\text{eff}}$ and the constant appearing in the potential
is proportional to $\xi\kappa^{2}\nu^{2}$, therefore, if $\nu^2
\sim 1/ \kappa^2 \xi$ as it happens in the exact solution we
found, the two constants are of the same order. This behaviour is
different from the exact MTZ solution where the potential
asymptotically becomes a cosmological constant which is the same
with the cosmological constant of the asymptotically AdS spacetime
seen in Eqs. (\ref{bhsol}), (\ref{kdd}). Hence, in our case, the
presence of the scalar field and its coupling to torsion modifies
the asymptotic form of the spacetime.

In the linearized large distance solutions we discuss in this Section the length scales,
defined by the cosmological constants discussed above, depend on two quantities, the
derivative coupling constant $\xi$ and the integration constant $\nu$ which expresses the
scalar charge. Thus, contrary to the previous wormhole solution, in this case the two
constants are not interrelated and this is explained since the linearized solution found
is a general asymptotic solution with the correct number of integration constants. Note
also that the derivative coupling constant $\xi$ enters the solution only through the
specific combination $\xi\kappa^{2}\nu^{2}$, while the integration constant $\nu$ appears
also independently. This combination is a remnant of the interaction term
$Tg^{\mu\nu}\partial_{\mu}\phi\partial_{\nu}\phi$ in the initial Lagrangian. Therefore,
the value of $\xi$ is not significant on its own, but only in combination with the
integration constant $\nu$ of the scalar field.

We note here that the aforementioned AdS behaviour, as opposed to
the dS behaviour, creates an (extra) attraction force at local
level at large distances, which may be useful at galactic or
cluster of galaxies scales, and in particular can offer a novel
mechanism for their stability. Also in the metric (\ref{duk}) the
term $R^{2-m}$ expresses another large distance scale where
gravity is modified, and for instance, in the case where $m=1$
this term becomes proportional to $R$. Then in
\cite{Mannheim:1988dj} it was claimed that   in Weyl gravity a
linear term $R$ can successfully explain the galactic rotation
curves without the need of adding dark matter.

%%%%%%%%%%%%%%%%%%%%%%%%%%%%%%%%%%%%%%%%%%%%%%%%%%%%%%%%%%%%%%%%%%%%%%%%%%%%

\item

{\underline{Case B}}

For $\sigma>\frac{1}{2}$ we have $\eta>0$. Then the only
consistent case is for $\varepsilon=-1$,
$\frac{1}{4}<\xi\kappa^{2}\nu^{2}<\frac{1}{2}$ (thus  the coupling
$\xi$ must be positive) and $\eta>\frac{\kappa^{2}}{\nu^{2}}$. The
values of $\lambda,\mu$ are $\mu>0$, $-2<\lambda<-3/2$, and thus
it should be $c_{\mu}=0$  otherwise $Y_1$ keeps increasing at
large distances and the condition $|Y_1| \ll |Y_0|$ is not
satisfied. Therefore, for $\sigma>\frac{1}{2}$ the solution becomes
\begin{equation}
Y=\frac{2\eta\nu^{2}}{3(2\sigma\!-\!1)}+c_{\ell}e^{-\ell
x}-\frac{2\sigma}{3\nu^{2}(2\sigma\!-\!1)} e^{-2x} \label{jca}
\end{equation}
with
\begin{eqnarray}
 \frac{3}{2}<\ell=2(1\!-\!\sigma)\!+\!\sqrt{1\!-\!2\sigma\!+\!4\sigma^{2}}<2\,,
\end{eqnarray}
or in terms of $R$
\begin{equation}
Y=\frac{2\eta\nu^{2}}{3(2\sigma\!-\!1)}+c_{\ell}R^{-\ell}-\frac{2\sigma}{3\nu^{2}
(2\sigma\!-\!1)}
R^{-2} \,, \label{bca}
\end{equation}
where $c_{\ell}$ is an integration constant. The metric
(\ref{jsb}) takes the form
\begin{equation}
ds^{2}=-N^{2}dt^{2}+\frac{dR^{2}}{\frac{2\eta\nu^{4}}{3(2\sigma-1)}R^{2}+\texttt{c}_{\ell}
R^{2-\ell}
-\frac{2\sigma}{3(2\sigma-1)}}+R^{2}d\Omega^{2}\,, \label{dun}
\end{equation}
where $\texttt{c}_{\ell}=\nu^{2}c_{\ell}$ is a redefined
integration constant. Following the same procedure as in in case A
we find similar forms for the lapse function and the scalar
field as in case A, while the form of the potential is found to be
\begin{equation}
V=\frac{(1\!+\!2\sigma)\nu^{2}}{2(1\!-\!2\sigma)}-\frac{3\!-\!m}{2\eta\nu^{2}}\frac{
\texttt{c}_{m}}{
R^{m}}
+\frac{3\!-\!8\sigma}{6(1\!-\!2\sigma)\eta\nu^{2}}\frac{1}{R^{2}}
+\frac{3\!-\!\ell}{2\eta\nu^{2}}\frac{\texttt{c}_{\ell}}{R^{\ell}}\,.
\label{ekg}
\end{equation}
The physical significance of this solution is similar to the
discussed case A solution.

\end{itemize}

\section{Conclusions}
\label{Conclusions}

In this work we investigated spherically symmetric solutions in
scalar-torsion gravity theories. Using the teleparallel
formulation of GR, which presents the torsion scalar field $T$,
we allowed a derivative coupling of a scalar field $\phi$
to the torsion scalar. The resulted theory
is a novel scalar-torsion theory different from the corresponding
curvature-based gravity theories. It has the advantage of giving
second order field equations leading to a healthy theory without
ghosts.

After extracting the general field equations, we applied them
in the background of a spherically symmetric geometry.
A novel feature of the arising coupled system
of gravity-scalar field equations is the appearance of an
off-diagonal equation, which couples the scalar field directly to a
metric function. The appearance of this equation constrains
further the system of the coupled field equations, making them
difficult to be solved analytically. In spite of that, we managed
to extract a single decoupled master equation, the solution of
which allows to find one by one all the other unknown functions.

Although the solution of the system of equations is unique for the metric
components, the scalar field profile, and the scalar-field potential, we have found
a specific form for the scalar potential which provides an exact solution
of the field equations. This solution is special from the point of view of
the number of integration constants, and as a result, both the integration
constant of the scalar field and the asymptotic value of the potential depend
on the derivative coupling constant. Interestingly, the
obtained solution is a wormhole-like one, however it extends
inside the ``throat'' of the wormhole up to the origin, while in a
usual wormhole the solution extends outside the ``throat'' up to the
``mouth'' or up to infinity. Moreover, the accompanied scalar
field remains regular both at the origin and at the ``throat''.
Hence, this solution belongs to a new class of wormhole-like
solutions which can have interesting physical features.

Using the master equation, we obtained linearized solutions that
are valid at large radial distances. Interestingly enough, we
showed that at large distances the space is AdS, which implies
that in the model at hand, at large distances we obtain a negative
effective cosmological constant. However, the asymptotic form of
the potential provides a different negative  constant. This is a
new feature, not present for example in the MTZ solution, and is
due to the derivative coupling along with the non-vanishing
asymptotic scalar field. Hence, in our case, the presence of the
scalar field and its coupling to torsion modifies the asymptotic
form of the spacetime. In these solutions the derivative coupling
constant and the integration constant of the scalar field are not
interrelated, contrary to the previous wormhole solution, and this
is due to the fact that the solutions are general from the point
of view of the number of integration constants.

The fact that the  derivative coupling creates an anti de-Sitter
space in the present solution, instead of a de-Sitter one, gives
an attractive term around a spherical object, which might have
interesting implications on structure formation and cosmology. To
the same direction may act the fact that apart from the above
behavior, the radial metric potential acquires another
contribution at smaller distances that can be close to
linear-in-distance, and thus playing the role of an effective dark
matter component.

In summary, as we observe, the  derivative coupling in the
framework of teleparallel gravity brings novel features that
cannot be obtained in the usual curvature nonminimal couplings.
However, a crucial test for the scalar-torsion theories to give
distinct features from the scalar-tensor theories of GR, is to
find exact solutions representing a central object, like a black
hole with scalar hair. Before considering this theory as a
candidate for the description of nature it has to be shown that it
is viable phenomenologically and the model parameters have to be
determined.

\begin{appendix}
\section{ Calculation of the master equation}

In this appendix we will give the technical details for obtaining
the master equation (\ref{jsitext}). Defining the variables
\begin{eqnarray}
&&Y\equiv y^{2}\label{Yy2}\\
&&Z\equiv z^{2}\label{Zz2}~,
\end{eqnarray}
 we find that  $\dot{y}=\frac{1}{2}\frac{dY}{dx}$,
$\dot{z}=\frac{1}{2}\frac{dZ}{dx}$ and therefore equations
(\ref{1ccf})-(\ref{3ccf}) become
\begin{eqnarray}
\frac{dY}{dx}+3Y-\frac{e^{-2x}}{\nu^{2}}
+\frac{\kappa^{2}(2V\!+\!\nu^{2})}{2\nu^{2}(1\!-\!2\xi\kappa^{2}\nu^{2})}=0~,\label{1ccc}
\\
YZ=F^{2}~,
\label{2ccc}\\
\frac{dZ}{dx}+Z-\frac{e^{-2x}}{\nu^{2}}
+\frac{3\kappa^{2}(2V\!+\!\nu^{2})}{2\nu^{2}(1\!-\!2\xi\kappa^{2}\nu^{2})}=0~.
\label{3ccc}
\end{eqnarray}
Equations (\ref{1ccc}), (\ref{3ccc}) are linear differential
equations, whose solutions are
\begin{eqnarray}
&&Y(x)=e^{-3x}\left(c_{1}+\frac{e^{x}}{\nu^{2}}-\frac{\eta}{3}\nu^{2}e^{3x}
-2\eta\int \!Ve^{3x}dx\right)~,\label{Y}\\
&&Z(x)=e^{-x}\left(c_{2}-\frac{e^{-x}}{\nu^{2}}-3\eta\nu^{2}e^{x}
-6\eta\int \!Ve^{x}dx\right)\,, \label{Z}
\end{eqnarray}
where $c_{1},c_{2}$ are integration constants and we have defined
the constant
\begin{equation}
\eta\equiv\frac{\kappa^{2}}{2\nu^{2}(1-2\xi\kappa^{2}\nu^{2})}\,.
\label{hud}
\end{equation}
Finally, inserting (\ref{Y}), (\ref{Z}) in (\ref{2ccc}) we obtain
an integro-differential equation for the potential $V$, considered
as a function of $x$, i.e. $V=V(x)$. Therefore, if the potential
$V(x)$ was given, in general the system would be inconsistent, and
the reason behind this is the extra off-diagonal equation
(\ref{4}). However, if the potential $V(x)$ satisfies this
integro-differential equation the system is consistent.

Let us transform the aforementioned integro-differential equation
into a differential equation. Defining the function
\begin{equation}
h(x)=\int V e^{3x}dx~, \label{hdp}
\end{equation}
we immediately obtain
\begin{equation}
V=e^{-3x}\frac{dh}{dx}~, \label{bdj}
\end{equation}
while $\int V e^{x}dx=\int e^{-2x}\frac{dh}{dx} dx=e^{-2x}h+2\int
e^{-2x} h \,dx$. Using $h(x)$, equations (\ref{Y}) and (\ref{Z})
give
\begin{eqnarray}
&&Y(x)=e^{-3x}\Big(c_{1}+\frac{e^{x}}{\nu^{2}}-\frac{\eta}{3}\nu^{2}e^{3x}
-2\eta h\Big)~,\label{Ya}\\
&&Z(x)=e^{-x}\left(c_{2}-\frac{e^{-x}}{\nu^{2}}-3\eta\nu^{2}e^{x}-6\eta
e^{-2x}h -12\eta\int \!e^{-2x}h\,dx\right)\,. \label{Za}
\end{eqnarray}
Thus, replacing $Z$ from (\ref{2ccc}) in (\ref{Za}) and
differentiating we get
\begin{equation}
\frac{d}{dx}\Big(e^{x}\frac{F^{2}}{Y}\Big)=\frac{e^{-x}}{\nu^{2}}
-3\eta\nu^{2}e^{x}-6\eta e^{-2x}\frac{dh}{dx}\,. \label{nyk}
\end{equation}
Additionally, from (\ref{2ccf}) we deduce that
\begin{equation}
F=\tilde{\eta}\Big(\frac{1}{2\nu^{2}\eta}e^{-2x}+\frac{\nu^{2}}{2}-e^{-3x}\frac{dh}{dx}
\Big)\,,
\label{koi}
\end{equation}
where we have introduced the constant
\begin{equation}
\tilde{\eta}\equiv\frac{\kappa^{2}}{\nu^{2}(1-6\xi\kappa^{2}\nu^{2})}\,.
\label{owk}
\end{equation}
Therefore, from (\ref{Ya}), (\ref{koi}), the function $F$ can be
expressed as
\begin{equation}
F=\frac{\tilde{\eta}}{2\eta}\Big(\frac{dY}{dx}+3Y+2\eta\nu^{2}\Big)\,,
\label{las}
\end{equation}
while (\ref{nyk}) becomes
\begin{equation}
\frac{d}{dx}\Big(e^{x}\frac{F^{2}}{Y}\Big)
=3e^{x}\frac{dY}{dx}+9e^{x}Y-\frac{2}{\nu^{2}}e^{-x}\,.
\label{jjf}
\end{equation}
Hence, given (\ref{las}), equation (\ref{jjf}) is a second order
autonomous equation for $Y(x)$ which is also written as
\begin{equation}
2F\frac{dF}{dx}-\Big(\frac{F^{2}}{Y}+3Y\Big)\frac{dY}{dx}+F^{2}-9Y^{2}+\frac{2}{\nu^{2}}e^
{-2x}Y=0\,
.
\label{sho}
\end{equation}
Finally, using (\ref{las}) and (\ref{sho}) we extract the
differential equation for $Y(x)$ as
\begin{equation}
2\frac{d^{2}Y}{dx^{2}}-\frac{1}{Y}
\Big(\frac{dY}{dx}\Big)^{2}+2\Big(2-\frac{\eta\nu^{2}}{Y}\Big)\frac{dY}{dx}
+3Y+2\eta\nu^{2} -12\frac{\eta^{2}}{\tilde{\eta}^{2}}Y
\frac{\frac{dY}{dx}+3Y-\frac{2}{3\nu^{2}}e^{-2x}}
{\frac{dY}{dx}+3Y+2\eta\nu^{2}}=0\,. \label{jsi}
\end{equation}

In summary, we have managed to replace the complicated system of
field equations (\ref{1})-(\ref{eqmsc}) with just the equation
(\ref{jsi}) plus decoupled equations.

\end{appendix}

\begin{acknowledgments}
The authors wish to thank V. Cardoso and F. Lobo for useful
discussions. E.P. is  partially supported by ARISTEIA II action of
the operational programme education and long life learning which
is co-funded by the European Union (European Social Fund) and
National Resources. The research of E.N.S. is implemented within
the framework of the Action ``Supporting Postdoctoral
Researchers'' of the Operational Program ``Education and Lifelong
Learning'' (Actions Beneficiary: General Secretariat for Research
and Technology), and is co-financed by the European Social Fund
(ESF) and the Greek State.
\end{acknowledgments}

\end{document}